\documentclass[journal,twoside,web]{ieeecolor}
\usepackage{tmi}
\usepackage{cite}
\usepackage{amsmath,amssymb,amsfonts}

\usepackage{mathrsfs}

\usepackage{algorithmic}
\usepackage{graphicx}
\usepackage{textcomp}
\def\BibTeX{{\rm B\kern-.05em{\sc i\kern-.025em b}\kern-.08em
    T\kern-.1667em\lower.7ex\hbox{E}\kern-.125emX}}
\begin{document}
\title{LoMAE: Low-level Vision Masked Autoencoders for Low-dose CT Denoising}
\author{Dayang Wang, \IEEEmembership{Student Member, IEEE}, Yongshun Xu, Shuo Han, Zhan Wu, \\
Li Zhou, Bahareh Morovati, Hengyong Yu*, \IEEEmembership{Senoir Member, IEEE}
\thanks{The paper was submitted on Oct. 1, 2023.  
This work was supported in part by NIH/NIBIB under grants R01EB032807 and R01EB3473, and NIH/NCI under grant R21CA264772.}
\thanks{D. Wang, Y. Xu, S. Han, L. Zhou, B. Morovati, and H. Yu are with the Department of Electrical and Computer Engineering, University of Massachusetts Lowell, Lowell, MA, USA, 01854.}
\thanks{Z. Wu is with the Laboratory of Image Science and Technology, Southeast University, Nanjing 210096, China, and also with the Key Laboratory of Computer Network and Information Integration, Southeast University, Ministry of Education, Nanjing 210096, China.}
\thanks{Dr. H. Yu is the corresponding author (e-mail: hengyong-yu@ieee.org).}
}

\maketitle

\begin{abstract}
Low-dose computed tomography (LDCT) offers reduced X-ray radiation exposure but at the cost of compromised image quality, characterized by increased noise and artifacts. 
Recently, transformer models emerged as a promising avenue to enhance LDCT image quality. However, the success of such models relies on a large amount of paired noisy and clean images, which are often scarce in clinical settings. In the fields of computer vision and natural language processing, masked autoencoders (MAE) have been recognized as an effective label-free self-pretraining method for transformers, due to their exceptional feature representation ability. However, the original pretraining and fine-tuning design fails to work in low-level vision tasks like denoising. In response to this challenge, we redesign the classical encoder-decoder learning model and facilitate a simple yet effective low-level vision MAE, referred to as LoMAE, tailored to address the LDCT denoising problem. Moreover, we introduce an MAE-GradCAM method to shed light on the latent learning mechanisms of the MAE/LoMAE. Additionally, we explore the LoMAE's robustness and generability across a variety of noise levels. Experiments results show that the proposed LoMAE can enhance the transformer's denoising performance and greatly relieve the dependence on the ground truth clean data. It also demonstrates remarkable robustness and generalizability over a spectrum of noise levels. 

\end{abstract}


\section{Introduction}
In recent years, Low-dose computed tomography (LDCT) has become the mainstream in clinical applications due to the potential risks from x-ray radiation in normal-dose CT (NDCT). Nonetheless, LDCT images exhibit compromised image quality and reduced diagnostic value, presenting a significant barrier for their widespread application. To overcome this issue, numerous deep learning models have been explored along this direction \cite{chen2017low,RN72,shan2019competitive,wang2022ctformer}.  


\begin{figure*}
\centering
\includegraphics[width=.8\linewidth]{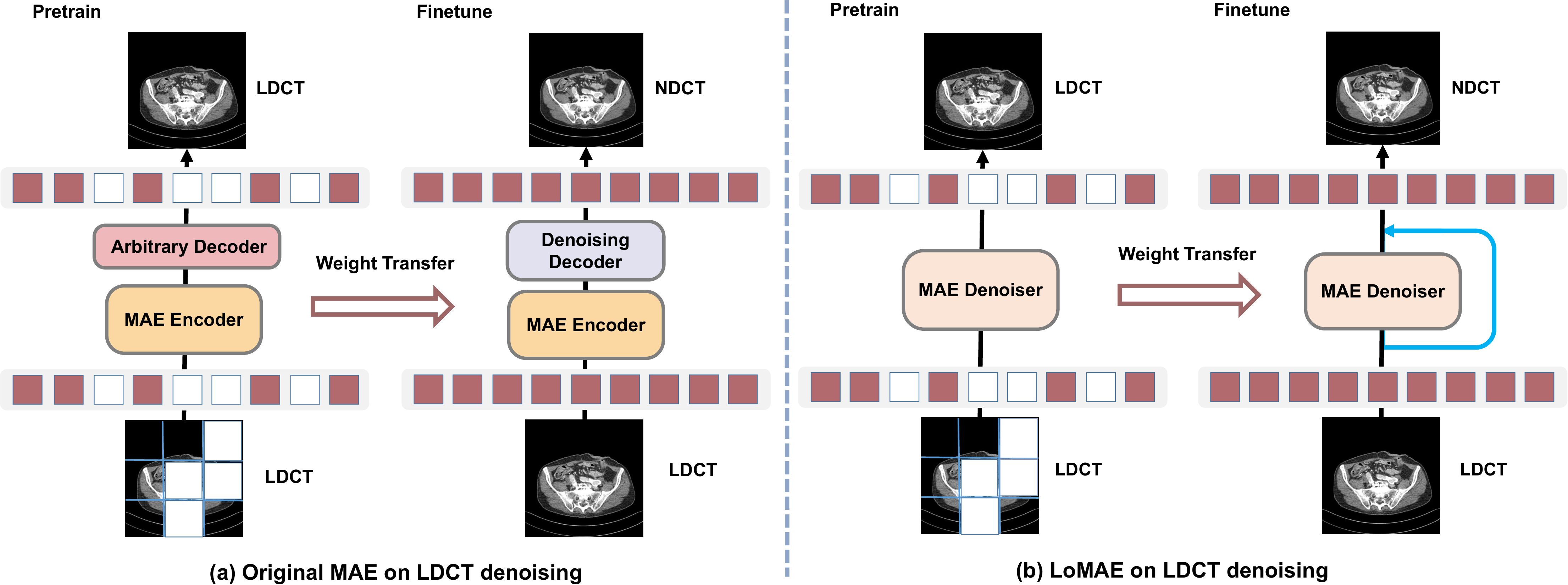}
\caption{Pipelines of the original MAE (a) and LoMAE (b) in LDCT denoising.  } \label{wholeflow1}
\end{figure*}

Recently, masked autoencoders (MAE) have emerged as an excellent self-supervised learning strategy for various computer vision tasks \cite{he2022masked,xie2022simmim}. 
He \textit{et. al.} revealed that small fraction of an image can infer complex and holistic visual concepts like semantics \cite{he2022masked}. Feichtenhofer \textit{et. al.} applied MAE on videos and found that the MAE method can almost eliminate inductive bias on spacetime with strong representations \cite{feichtenhofer2022masked}. Germain \textit{et. al.} modified MAE to data distribution estimation and interpreted the MAE output as conditional probabilities and the full joint probability, thus prototyping a strong distribution estimator with faster speed and higher scalability \cite{germain2015made}. Gao \textit{et. al.} showed that more discriminative representations can be further learned via a MAE multi-scale hybrid convolution-transformer architecture \cite{gao2022convmae}. Furthermore, they proposed a block-wise masking strategy to prevent information leakage and guarantee computational efficiency. Besides these explorations on deep architecture design and general natural images, the MAE was also explored in the medical imaging field. Zhou et al. were among the first to demonstrate that MAE can improve the performance of medical image segmentation and classification \cite{zhou2022self}. However, these works are on high-level vision tasks. While masked image modeling has been explored as a robust denoiser \cite{tihon2021daema,wu2022denoising,zhao2023self}, currently, there lacks sufficient exploration of MAE as self-pretraining on low-level vision tasks like LDCT denoising. 

Hence, we are inspired to delve into the potential of the MAE in LDCT denoising. We believe this study is significant for two compelling reasons: i) clinical applications often face the challenge of limited ground truth data availability. The self-pretraining paradigm of the MAE can reap the benefits of the unlabeled data, making it an ideal choice for LDCT denoising task. ii) LDCT denoising places significant importance on structural preservation and enhancement. But, the anatomical structures within a CT image are supposed to be intricately connected both mechanically and functionally. Here, the MAE's innate ability to aggregate contextual information and infer masked structures has great potential to fortify the interdependence between anatomical regions, thus complementing structural loss. Consequently, not only does this reinforcement bolster the model's robustness, but it also potentially augments its capacity to generalize across a diverse spectrum of noise types and levels, rendering it invaluable in LDCT denoising endeavors.



The contributions of this paper are fourfolds.
1) Tailored low-level vision MAE: a low-level vision masked autoencoder (LoMAE) framework is designed and specifically tailored to the demands of LDCT denoising tasks. This novel approach ensures that our method is well-aligned with the unique challenges presented by LDCT image enhancement.
2) Reduced ground truth dependency: our LoMAE represents a breakthrough in reducing the reliance on ground truth data. Remarkably, even with access to just two patients' clean data, LoMAE outperforms fully supervised baseline models. This achievement underscores the practicality and effectiveness of LoMAE in scenarios where annotated data is scarce.
3) Robustness and generalizability: the exceptional robustness and generalizability of the LoMAE model are demonstrated across a spectrum of noise levels. This finding underscores the model's resilience and its ability to deliver consistent performance in different real-world settings.
4) MAE-GradCAM insight: for the first time, the MAE-GradCAM method is introduced, shedding light on the intricate pretraining process of the Masked Autoencoder. This novel approach unveils hidden patterns and provides valuable insights into the inner workings of MAE-based models.


This manuscript is a significant extension of the IEEE International Symposium on Biomedical Imaging (ISBI) 2023 paper entitled "\textit{Masked Autoencoders for Low-dose CT Denoising}" \cite{wang2023masked,wang2022masked}, which is the first work to adapt MAE to the low-level vision tasks like denoising. In this paper, more extensive experiments are conducted to demonstrate the properties of the proposed LoMAE strategy on the LDCT denoising problem. 

\section{Related Works}

There are two lines of research which are highly related to this manuscript on LDCT denoising. 

\textbf{Convolutional Models.}
In the past few years, deep learning has been prevailing in various science and engineering fields \cite{fan2021sparse,jia2021nondestructive,morovati2023reduced}. Its applications in LDCT have also achieved many state-of-the-art performances. Chen \textit{et al.} are the pioneers in designing a residual encoder-decoder convolutional model (RED-CNN) for the LDCT denoising \cite{RN69}. Shan proposed a modularized deep neural network (MAPNN) for LDCT reconstruction and achieved better qualities than commercial iterative reconstruction methods with higher structural fidelity and faster speed \cite{shan2019competitive}.  Fan \textit{et al.} introduced a quadratic encoder-decoder LDCT denoising network with more efficiency and robustness. Despite these achievements, the CNN models are limited in capturing global contextual information, thus less powerful to model long-stride structure dependence among different anatomical regions in the CT images.



\textbf{Transformer Models.} In recent years, transformer models have gained significant attention for their proficiency in capturing global contextual information. Several researches also extended their applications to the realm of LDCT denoising. Wang \textit{et al.} advanced the field by introducing a convolution-free encoder-decoder transformer, , providing insights into the static and dynamic latent learning behaviors of the model through attention map analysis \cite{wang2022ctformer}. Meanwhile, Yang et al. pioneered the development of a sinogram inner-structure transformer, leveraging sinogram inner-structures to suppress LDCT noise effectively \cite{yang2022low}. Recently, the Swin transformer has gained prominence as a versatile backbone architecture for a variety of downstream tasks \cite{RN74}. SwinIR is its notable variant, specifically tailored for tasks such as image denoising, super-resolution, and artifact reduction \cite{liang2021swinir}. Notably, some SwinIR-based transformer models or modules have also been introduced to CT image enhancement/denoising and demonstrating exceptional performances \cite{puttaguntaa2022swinir}. Fan \textit{et al.} also proposed another U-shaped Denoising transformer, namely SUNet \cite{fan2022sunet}, which was based on Swin transformer modules and achieved the SOTA performance in image denoising \cite{fan2022sunet}. 


\section{Methods}

\subsection{Noise Model}

In LDCT images, two primary types of noise are prevalent. The first type is thermal noise, which emanates from the detection system. This noise adheres to a Gaussian distribution and is structurally independent.
The second type is photon statistical noise due to the inherent quantum fluctuations during the X-ray emission. This noise typically conforms to a Poisson distribution that is associated with the object structures. The LDCT noise is a complex interplay of both Gaussian and Poisson components, as expressed by the equation: 
\begin{equation}
    \boldsymbol{\psi} \thicksim  \mathcal{N}(0,\sigma^2) +  \mathcal{P} (\xi(\boldsymbol{\rho})),
\end{equation}
where $\boldsymbol{\psi}$ is the noise variable, $\mathcal{N}(0,\sigma^2)$ denotes the Gaussian distribution whose variance is $\sigma^2$, and $\mathcal{P} (\xi(\boldsymbol{\rho}))$ is Poisson distribution in which  $\xi(\boldsymbol{\rho})$ represents the latent mapping from projections $\boldsymbol{\rho}$ of the clean image structure to the distribution expectancy.

The proposed LDCT denoising model directly learns from paired noisy LDCT images to their corresponding clean NDCT images $\boldsymbol{y}$ directly. Experimentally, $\mathbf{L}_1$ and $\mathbf{SSIM}$ losses are employed to suppress these noises during training. Mathematically, the $\mathbf{L}_1$ and $\mathbf{SSIM}$ losses are calculated as follows: 


\begin{equation}
    \mathbf{L}_1( \tilde{\boldsymbol{y}},\boldsymbol{y})=||\tilde{\boldsymbol{y}}-\boldsymbol{y} ||_1,
\end{equation}
\begin{equation}
    \mathrm{SSIM}({\tilde{\boldsymbol{y}}},{\boldsymbol{y}}) = \frac{(2u_{\tilde{\boldsymbol{y}}} u_{\boldsymbol{y}} + c_1)(2\sigma_{{\tilde{\boldsymbol{y}}}{\boldsymbol{y}}}+c_2)}{(u_{\tilde{\boldsymbol{y}}}^2+u_{\boldsymbol{y}}^2+c_1)(\sigma_{\tilde{\boldsymbol{y}}}^2+\sigma_{\boldsymbol{y}}^2+c_2)},
\label{ssim}
\end{equation}
where $\tilde{\boldsymbol{y}}$ denote the prediction. $u_a$ and $\sigma_a^2$ represent the mean and variance of variable $a$, respectively.
$c_1$ and $c_2$ are two small values to avoid zero denominator. The final loss is the combination of the two losses:
\begin{equation}
    \mathcal{L}_f({\tilde{\boldsymbol{y}}},{\boldsymbol{y}}) = \alpha \cdot \mathbf{L}_1( \tilde{\boldsymbol{y}},\boldsymbol{y}) + \beta \cdot \mathbf{SSIM}(\tilde{\boldsymbol{y}},\boldsymbol{y}),
    \label{losse}
\end{equation}
where $\alpha$ and $\beta$ are hyperparameters to balance the losses. The $\mathbf{L}_1$ loss serves to suppress structural-independent Gaussian noise while preserving the image's brightness and color fidelity. Concurrently, the $\mathbf{SSIM}$ loss plays a pivotal role in mitigating Poisson noise by maintaining contrast in the high-frequency regions, thereby enhancing crucial structural features for accurate denoising. 







\subsection{Low-level Masked Autoencoder Design}
The model flowchart in Figure \ref{wholeflow1} demonstrates the straightforward implementation of MAE learning paradigm for LDCT denoising. In the self-pretraining, the LDCT images are subjected to patch-wise masking at a high rate, typically $75\%$ as proposed in previous work \cite{he2022masked}. These masked images are then processed through an encoder and an arbitrary decoder to reconstruct the original image. In fine-tuning, the lightweight decoder is replaced by a specialized decoder tailored to match the target task, such as LDCT denoising in our context. While this approach demonstrates its applicability in various high-level vision tasks such as classification, segmentation, and object detection, its effectiveness diminishes in the context of low-level vision tasks like denoising.


To address this challenge, we introduce a novel masked image-based self-pretraining approach, termed LoMAE, tailored for low-level vision tasks like LDCT denoising. Unlike conventional methods, LoMAE (illustrated in Figure \ref{wholeflow1}(b)) adopts a more streamlined architecture, consolidating the encoder and decoder into a single MAE denoiser. This transformation is grounded in the shared image-to-image training paradigm between LDCT denoising and MAE pretraining. However, it is essential to emphasize the fundamental distinction between these two tasks. Particularly, MAE pretraining focuses on supplementing and restoring missing structures while the denoiser primarily aims at Gaussian/Poisson noise removal. To bridge this gap and enhance structural preservation, we redesign the pretraining model by adding a front-to-end residual shortcut within the pretraining model. This redesign empowers the model to better tackle denoising tasks while maintaining information consistency and avoiding structural loss.


Particularly, in the pertaining stage, the model learns from masked LDCT to whole LDCT images $\boldsymbol{x}$,  
\begin{equation}
   \underset{\mathbf{W_o}}{ \min} \ \  \mathcal{J}_p(\mathbf{W_o};\boldsymbol{x})=\mathbf{L}_1 (\mathcal{D}^-(\mathbf{W_o};\boldsymbol{x} \cdot \boldsymbol{\mathcal{M}}),\boldsymbol{x}),
    \label{equ1}
\end{equation}
where $\mathcal{D}^-(\mathbf{W_o};\boldsymbol{x})$ is the denoising model without front-to-end shortcut, and $\mathbf{W_o}$ is a collection of initial parameters before pretraining. After well-trained, let's assume the parameters become $\mathbf{W_p}$. $\boldsymbol{\mathcal{M}}$ represents the mask on the patch-wise tokens from the images. 


During finetuning, the model is trained from LDCT to NDCT image $\boldsymbol{y}$ to learn the noise removal process. Here, we remove the image masks and transfer the model weights from the pretraining stage, maintaining an identical architecture. This finetuning stage can be mathematically expressed as:

\begin{equation}
   \underset{\mathbf{W_p}}{ \min} \ \  \mathcal{J}_f(\mathbf{W_p};\boldsymbol{x}) =
    \mathcal{L}_f (\mathcal{D}^-(\mathbf{W_p};\boldsymbol{x})+\boldsymbol{x},\boldsymbol{y}). 
    \label{equ2}
\end{equation}

\begin{figure}
\centering
\includegraphics[width=\linewidth]{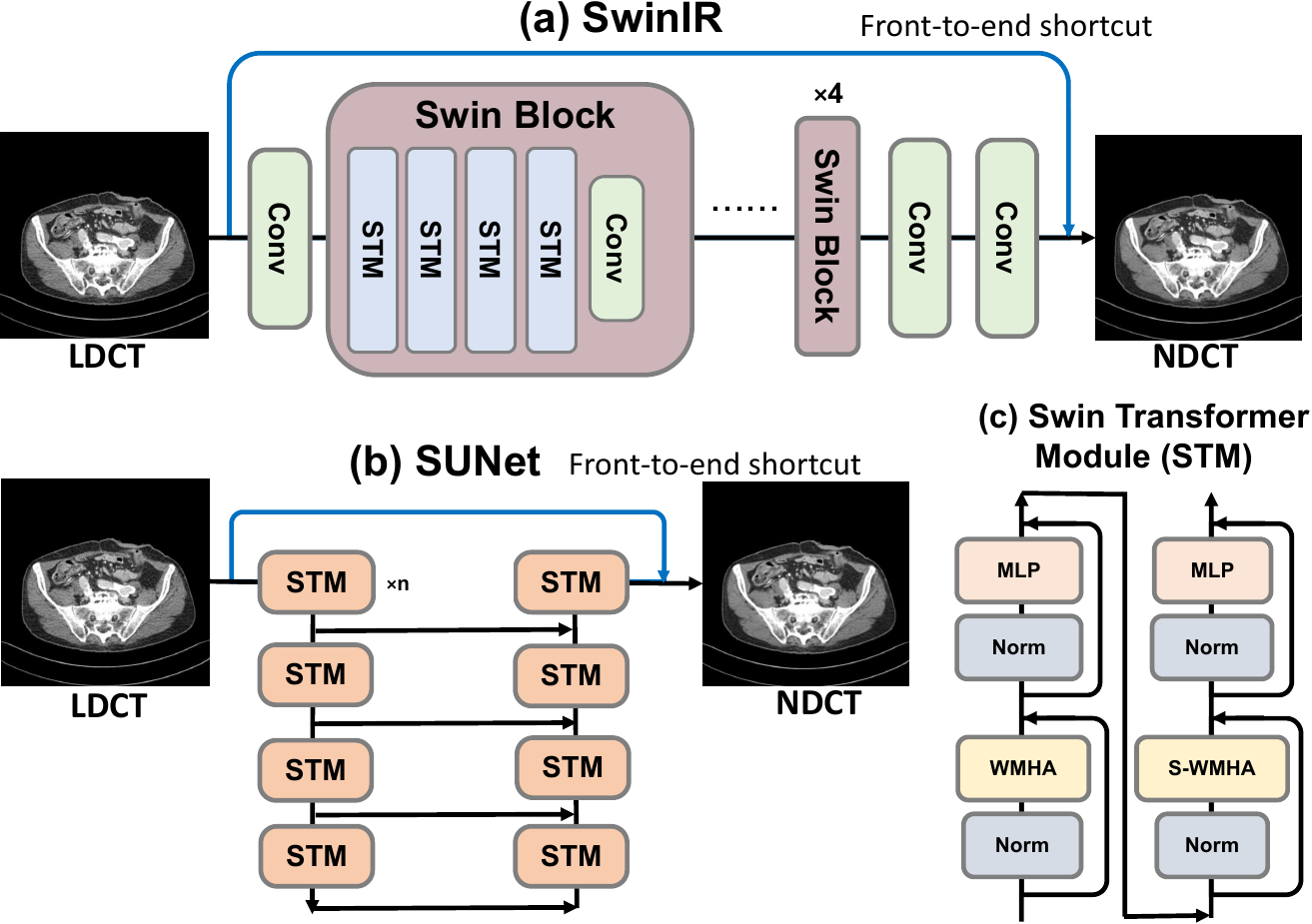}
\caption{The architecture of the SwinIR and SUNet as MAE denoiser. 
Front-to-end shortcuts are removed in the MAE pretraining stage and then are reconnected in the finetuning stage.}
\label{swin}
\end{figure}

\subsection{Swin Transformer}
The Swin transformer introduced a hierarchical attention mechanism featuring shifted windows, enhancing contextual information fusion while controlling computational costs. Initially, the image is divided into non-overlapping patches, followed by linear mapping into tokens as transformer block input. In a Swin transformer module (Figure \ref{swin}(c)), input token sequence $\mathbf{z}^{l-1} \in R^{b \times w \times n \times d_o}$ first undergoes initial layer normalization (LN). Here, $b$ is batch size, $w$ is window count, $n$ is token count, and $d_o$ represents token embedding dimension. Then, $\mathbf{z}^{l-1}$ goes through linear transformations to derive query (Q), key (K), and value (V) components to compute the window-based multi-head attention (WMHA):

\begin{equation}
    \mathrm{WMHA}(\mathbf{Q},\mathbf{K},\mathbf{V}) = \mathrm{softmax}(\frac{\mathbf{Q}\mathbf{K^\top}}{\sqrt{d_k}}+\mathbf{B}) \mathbf{V},
\end{equation}
where $\frac{1}{\sqrt{{{d}_{k}}}}$ is the scaling factor. $\mathbf{B}$ is a bias matrix. 
Subsequently, an additional LN layer, a multiple-layer perceptron (MLP), and two residual shortcuts are employed to enhance feature inference.
As shown in Fig. \ref{swin} (c), a successive Swin transformer module is characterized as
\begin{equation}
\begin{cases}
   &  \hat{\mathbf{z}}^l = \mathrm{WMHA}(\mathrm{LN}(\mathbf{z}^{l-1})) + \mathbf{z}^{l-1} \\
   &  \mathbf{z}^l = \mathrm{MLP}(\mathrm{LN}(\hat{\mathbf{z}}^l)) + \hat{\mathbf{z}}^l \\
   &  \hat{\mathbf{z}}^{l+1} = \mathrm{S\text{-}WMHA}(\mathrm{LN}(\mathbf{z}^{l})) + \mathbf{z}^{l} \\
   &  \mathbf{z}^{l+1} = \mathrm{MLP}(\mathrm{LN}(\hat{\mathbf{z}}^{l+1})) + \hat{\mathbf{z}}^{l+1}. \\
\end{cases}
\end{equation}
Compared with the $\mathrm{WMHA}(\cdot)$ function, $\mathrm{S\text{-}WMHA}(\cdot)$ incorporates an extra cyclic shift operation to further enhance rich spatial information fusion. In addition, to reduce computational complexity and support hierarchical representation, the patch merging layers are interposed between successive transformer blocks to reduce the feature map size by half. 

In recent years, an abundance of deep learning models has harnessed Swin transformer modules as their foundation. Similarly, the transformer models we utilize to study the performance of the LoAME are also built upon Swin modules, namely SwinIR \cite{liang2021swinir} and SUNet \cite{fan2022sunet}, which represent two distinct denoising architectures, as shown in Fig. \ref{swin}(a) and (b). Both variants have demonstrated formidable performance in addressing low-level denoising challenges. 


\subsection{MAE Interpretation}
While the MAE has been effective across diverse areas, interpreting its inner workings remains a formidable challenging task as the majority of other deep models. Naturally, our curiosity extends to understand how the model complements the missing patches in the pretraining stage, thus facilitating robust structural preservation. Among the handful methods for peering into the workings of deep learning models, GradCAM \cite{selvaraju2017grad} stands out as a favored choice since it provides easy-to-understand saliency maps to elucidate what element models are relying on to make a decision. GradCAM has found widespread utility in a multitude of general computer vision tasks, spanning classification, regression, segmentation, and more. Motivated by this, we have endeavored to adapt the GradCAM methodology, here termed as MAE-GradCAM, to decode the latent learning behavior of MAE. Please note that this method can be applied to both the proposed LoMAE and original MAE since they share the same pretraining procedure. Specifically, the traditional GradCAM generates the saliency map $\boldsymbol{\mathcal{S}}$ through the following formula:
\begin{equation}
\begin{cases}
   & \boldsymbol{\mathcal{S}} = \mathrm{ReLU}(\sum_{k} \boldsymbol{\mathcal{G}}^k_c \mathbf{\mathcal{A}}^k) \\
   & \boldsymbol{\mathcal{G}}^k_c = \frac{1}{\mathrm{N}} \sum_{u,v} \frac{\partial y^c}{\partial \mathbf{\mathcal{A}}^k_{uv}}, 
\end{cases}
\end{equation}
where $y^c$ is the logits for the selected class. The selected layer is typically the last convolution layer prior to the classification head. However, these do not work in MAE because there is no concept of 'logits' in MAE and the last layer aims to restore the original pixels thus representing low-level features with detailed structures. 

To unravel the concealed mechanisms governing how MAE can fill the masked patches, the MAE-GradCAM is distinct from the original GradCAM in two crucial aspects. First, $\partial y^c$ is replaced by $\partial \mathbf{L_1}(\hat{y}_{p},y_{p})$, where $\hat{y}_{(i,j)\in p} \subset \hat{y}$ and $y_{(i,j)\in p} \subset y$ are prediction and target pixels within certain interested patches $p$. In this sense, the gradients are calculated within a defined region, allowing the MAE-GradCAM to unveil which areas positively contribute to the generation of the missing patches. Second, we choose the middle layer of the feature maps to visualize the gradients. In contrast to classification models where the last convolution usually encodes rich high-level semantic information, the last layer in the MAE training typically encapsulates low-level features devoid of semantic content. Consequently, we select the middle layer $\mathbf{\mathcal{A}}_m$ of base network as our area of interest since it includes both high-level features and spatial information. Therefore, the saliency map $\boldsymbol{\mathcal{S}}$ generated by MAE-GradCAM is defined as follows:
\begin{equation}
\begin{cases}
   & \boldsymbol{\mathcal{S}} = \mathrm{ReLU}(\sum_{k} \boldsymbol{\mathcal{G}}^k\mathbf{\mathcal{A}}^k_m) \\
   & \boldsymbol{\mathcal{G}}^k = \frac{1}{\mathrm{N}} \sum_{u,v} \frac{\partial \mathbf{L_1}(\hat{y}_{p},y_{p})}{\partial \mathbf{\mathcal{A}}^k_{uv}}.
\end{cases}
\end{equation}


\section{Experiments and Results}


\subsection{Dataset}

\textit{Low-dose dataset}: We employed the publicly available dataset from the \textit{2016 NIH-AAPM-Mayo Clinic LDCT Grand Challenge}\footnote{https://www.aapm.org/GrandChallenge/LowDoseCT/} \cite{RN83} for both model training and evaluation. This dataset encompasses 2,378 CT images with a slice thickness of 3.0 mm, collected from ten distinct patients. To ensure robustness, we implemented a ten-fold cross-validation strategy, where each fold reserved one patient for testing and the remaining for training. Due to GPU limitations, the original $512\times512$ images were uniformly resized to $256\times256$ using \textit{inter\_area} interpolation. Furthermore, techniques such as image rotation and flipping are utilized for data augmentation.


\textit{Simulated dataset}: To test the robustness and generalizability of the proposed model, we simulate multiple LDCT scans with different dose level based on \cite{RN83}. To mimic the realistic clinical environment, we first obtain the projection datasets of Radon transform from the Manifold and Graph Integrative Convolutional Network (MAGIC) \cite{xia2021magic}. Then, the CT images of different dose level are synthesized by adding Poisson and electronic noise to the clean projections. Here, the geometry of the scan is listed as follows: The distance from the x-ray source to the isocenter of image is 595 mm. The distance from the detector to the source is 1085.6 mm. The size of the image pixel is 0.6641 mm. The number of detector units is 1024, with a unit pitch of 1.2854 mm. With the obtained noise-free projections, the noisy projections are obtained by the following formula:
\begin{equation}
    \boldsymbol{y} = \mathrm{ln} \frac{\mathrm{I}_0}{\mathrm{Poisson}(\mathrm{I}_0 \mathrm{exp}(-\hat{\boldsymbol{y}})) + \mathrm{Normal}(0,\sigma_e^2)},
\end{equation}
where $\mathrm{I}_0$ represents the x-ray intensity. $\hat{y}$ is the noise-free sinogram and $\sigma_e^2$ is the variance in the normal distribution. In our experiment, the electronic noise variance is set to 10 according to \cite{xia2021ct,niu2014sparse}. Then, the noisy CT images are reconstructed using filtered backprojection algorithm in MAGIC.

\subsection{Experiment Settings}
Our experiments are conducted on Ubuntu 18.04.5 LTS system equipped with an Intel(R) Core(TM) i9-9920X CPU @ 3.50GHz. All models are implemented with PyTorch 1.5.0 and CUDA 10.2.0 on four NVIDIA 2080TI 11G GPUs for accelerated computations. Here are the experiment settings: 

\begin{itemize}
    \item The masked patch size is 8 with a masked rate of $75\%$ for both SwinIR and SUNet during  pretraining. 
    \item The SwinIR model was configured with block lengths of [4,4,4,4] with a constant embedding dimension of 60 and 6 heads for the multiple head attention. 
    \item The model depths for the SUNet are [2,2,2,2] for both encoder and decode with a uniform embedding dimension of 80 and 8 heads for attention. 
    \item In both the LoMAE pretraining and finetuning phases, we optimize the model over 50 epochs using the Adam optimizer. We initiate the learning rate at $1.5 \times 10^{-4}$ and implement a scheduled learning rate decay of 0.5 every 3,000 iterations.
    \item For both LoMAE pretraining and finetuning processes, the batch size is set to 1.  
\end{itemize}




\subsection{Result Comparison}
To assess the effectiveness of LoMAE pretraining for enhancing transformer denoising model, the following state-of-the-art models are investigated: REDCNN \cite{chen2017low}, MAPNN \cite{shan2019competitive}, SUNet \cite{fan2022sunet}, and SwinIR \cite{liang2021swinir}. Specifically, all these methods are rigorously implemented and optimized according to the officially disclosed codebases. Furthermore, we explore the integration of LoMAE as a pretraining strategy into SwinIR (SwinIR+LoMAE) and SUNet (SUNet+LoMAE). 

\begin{figure}
\centering
\includegraphics[width=\linewidth]{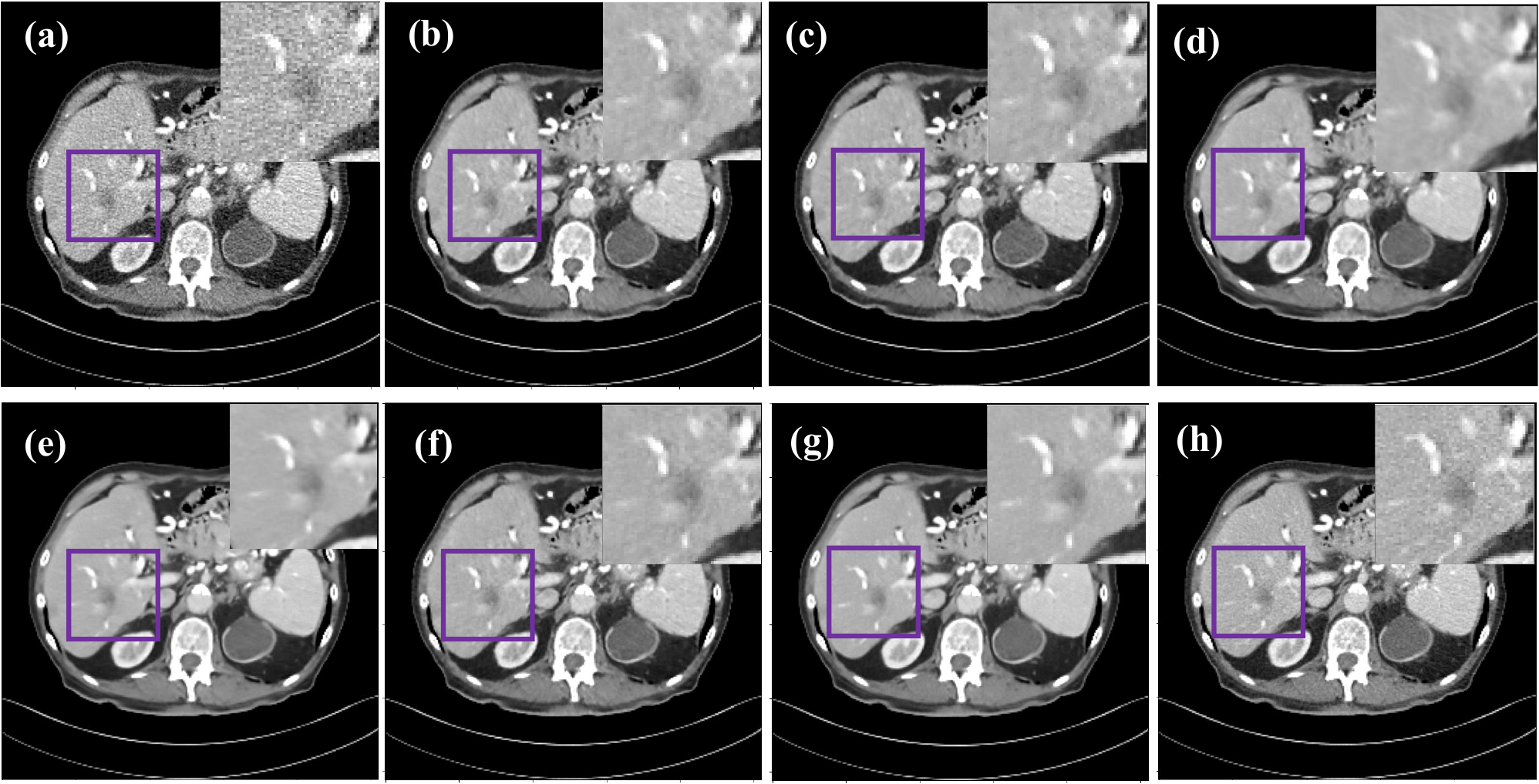}
\caption{The denoised results of Lesion 575 from patient L506. (a) LDCT, (b) REDCNN, (c) MAPNN, (d) SUNet, (e) SUNet+LoMAE, (f) SwinIR, (g) SwinIR+LoMAE, (h) NDCT. The display window is [-160,240] \textit{HU}.} 
\label{86whole}
\end{figure}

\begin{figure}
\centering
\includegraphics[width=\linewidth]{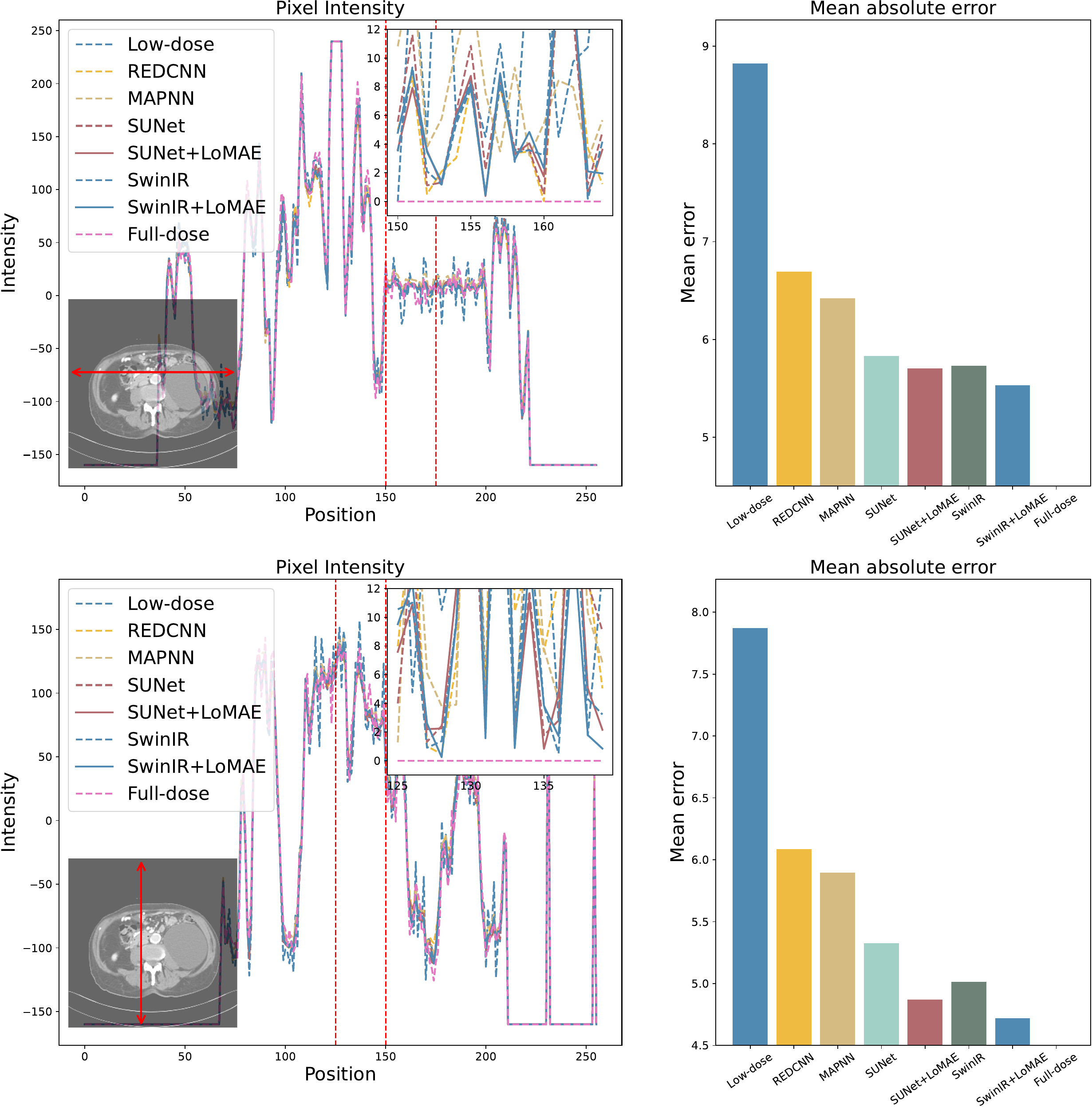}
\caption{The intensity profiles of different methods and the corresponding mean absolute errors on a representative slice. Notably, the left two figures feature a zoomed-in view that indicates the pixel-wise mean absolute error when compared to NDCT within the region of interest. } 
\label{intensity}
\end{figure}

\begin{figure}
\centering
\includegraphics[width=\linewidth]{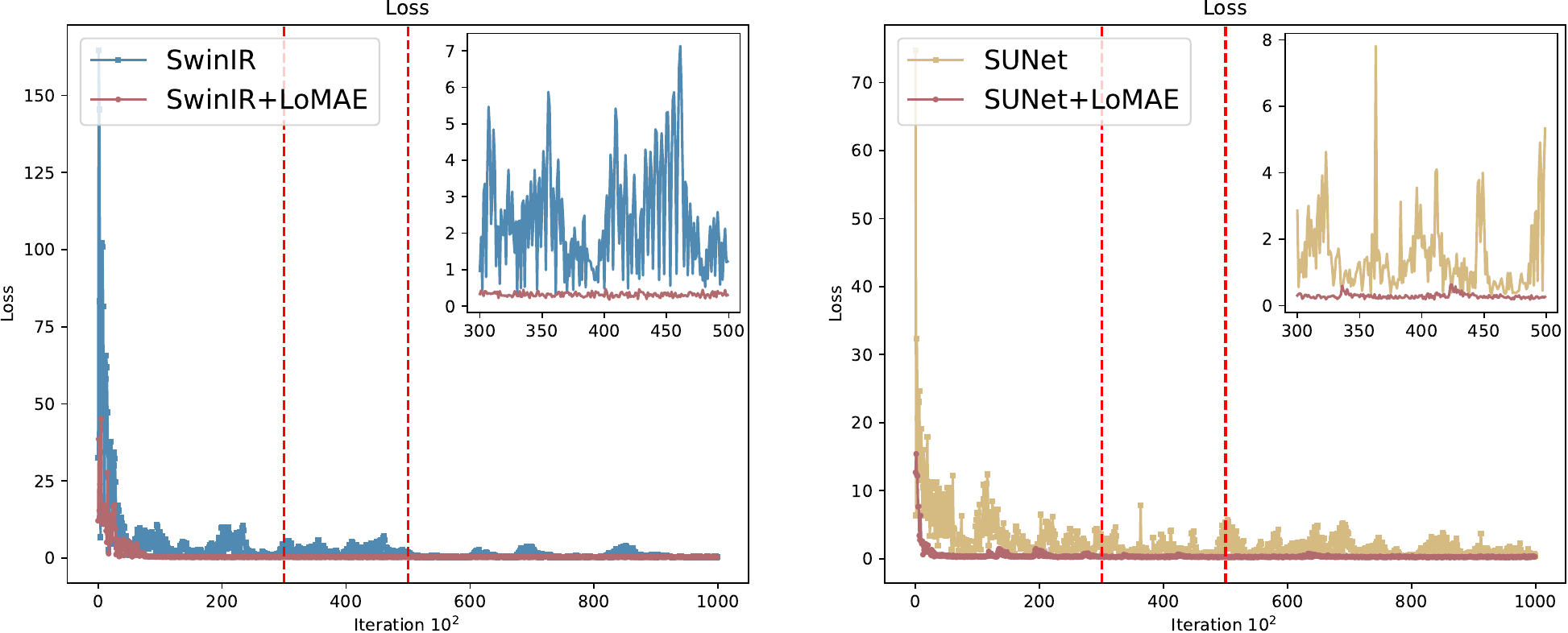}
\caption{The loss curves of the SwinIR and SUNet with their corresponding LoMAE methods.} 
\label{loss}
\end{figure}

Results in Fig. \ref{86whole} show that all these methods are effective in removing the noises from LDCT images. The SwinIR and SUNet can generate clearer noise-free images than the REDCNN and MAPNN. Upon closer examination of the regions of interest (ROIs) in Fig. \ref{86whole}, it's evident that our proposed approaches, SwinIR+LoMAE and SUNet+LoMAE, exhibit superior performance in generating more accurate and smoother lesion structures while retaining essential details when compared to the original models.
Moving on to Fig. \ref{intensity}, we provide visualizations of horizontal and vertical intensity profiles for a representative slice. Results indicate that the overall trend of the intensity profiles of the studied methods are close, whereas the proposed methods (SwinIR+LoMAE and SUNet+LoMAE) generate the most accurate structures, closely aligning with the NDCT image. Furthermore, the zoomed-in view and bar plot in Fig. \ref{intensity} highlight that our models exhibit the lowest mean absolute error in terms of intensities compared to others. In addition, as illustrated in Fig. \ref{loss}, the loss curves of the SwinIR+LoMAE and SUNet+LoMAE display a smoother trajectory compared to SwinIR and SUNet, respectively. This indicates that the LoMAE methods benefit from a much more stable and consistent training process than the original approaches. 

In addition to qualitative assessments, our model undergoes quantitative evaluation using two metrics including \textit{structural similarity} (SSIM) and \textit{root mean square error} (RMSE). Results in Tab. \ref{tab:quant} indicate that the transformer models (SwinIR and SUNet) can achieve better performance than the convolutional models (REDCNN and MAPNN). Notably, SwinIR achieves the highest scores in both metrics. Furthermore, the incorporation of LoMAE into SwinIR and SUNet leads to substantial quantitative improvements. Specifically, we observe an increase of 0.0063 and 0.0046 in SSIM, as well as reductions of 0.4877 and 0.4795 in RMSE for SwinIR+LoMAE and SUNet+LoMAE, respectively. In summary, our proposed LoMAE significantly enhances the denoising performance of transformer-based models, both qualitatively and quantitatively.

\begin{table}[h]
\caption{Quantitative results of different methods on L$506$. The bold-faced numbers are the best results. }
    \centering
    \scalebox{1.0}{
    \begin{tabular}{|l|c|c|}
    \hline
    Method  &  SSIM$\uparrow$  &   RMSE$\downarrow$   \\
    \hline
    LDCT      &0.9359  & 10.4833 \\
    REDCNN     &0.9501$\pm{0.0012}$  & 7.8016$\pm{0.0834}$ \\
    MAPNN      &0.9504$\pm{0.0019}$  & 7.5357$\pm{0.1342}$ \\
    \hline
    SUNet & 0.9555$\pm{0.0010}$  & 7.3147$\pm{0.1084}$\\
    SUNet+LoMAE* & 0.9601$\pm{0.0009}$  & 6.8352$\pm{0.1022}$\\
    \hline
    SwinIR   & 0.9546$\pm{0.0011}$ & 7.2232$\pm{0.0981}$\\
    SwinIR+LoMAE*  & \textbf{0.9609$\pm{0.0007}$} & \textbf{6.7355$\pm{0.0800}$}\\
    \hline
    \end{tabular}}
  \label{tab:quant}
    \hfill
\end{table}


\begin{figure}
\centering
\includegraphics[width=\linewidth]{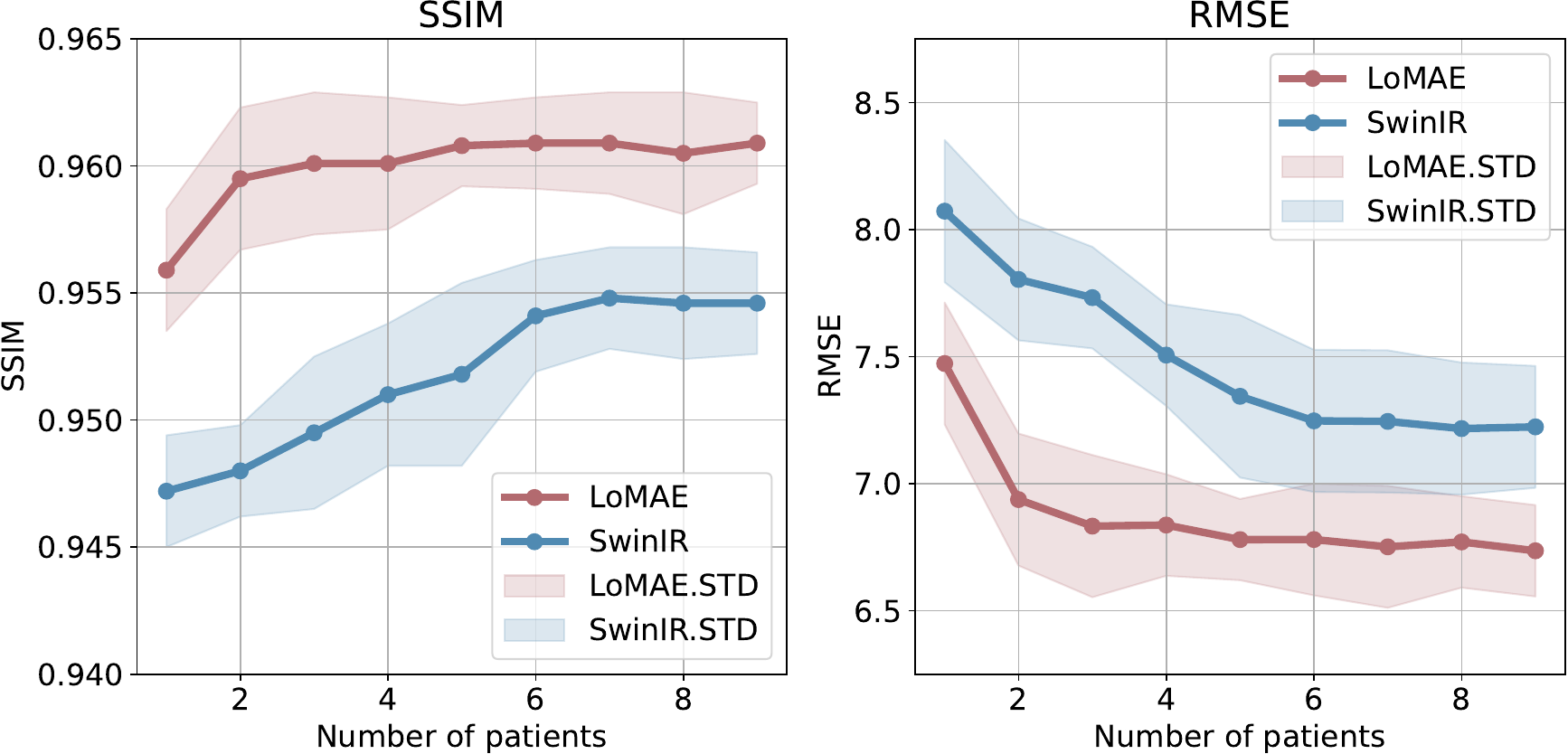}
\caption{The quantitative results of SSIM and RMSE using different numbers of patients for training.} 
\label{depend}
\end{figure}

\begin{figure*}
\centering
\includegraphics[width=\linewidth]{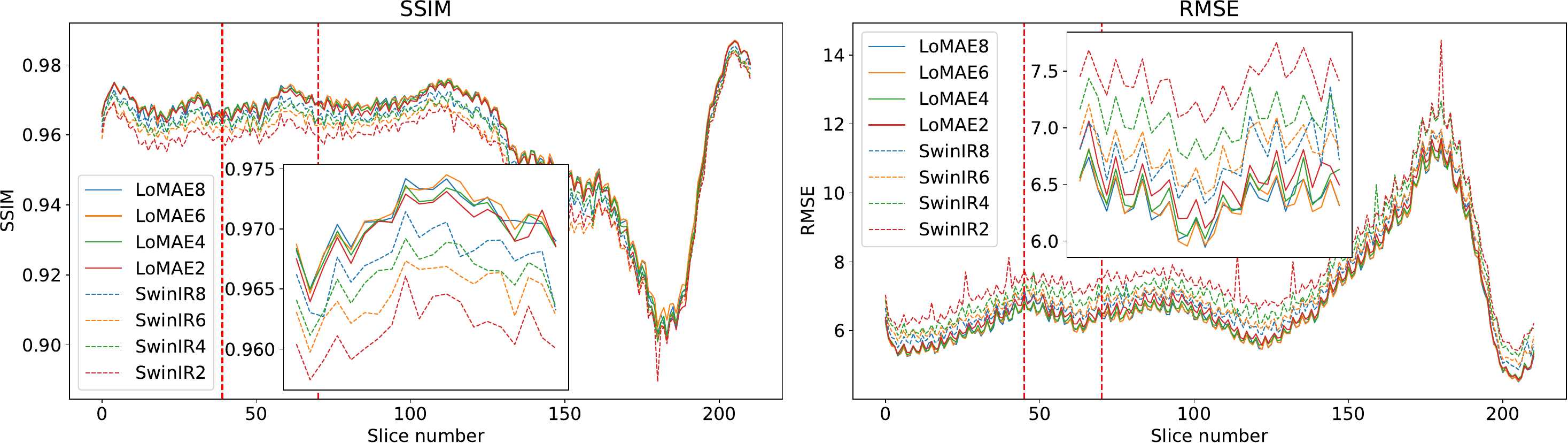}
\caption{The quantitative results of different methods on all testing slices from patient L506. Please note that the solid line '-' indicates the SwinIR+LoMAE method, while the dashed line '- -' denotes SwinIR.  } 
\label{scatter}
\end{figure*}

\subsection{Ground Truth Data Dependency}
In clinical applications, obtaining well-organized paired data for supervised deep models can be exceedingly challenging, if not impossible. This difficulty arises from the impracticality and ethical concerns associated with subjecting patients to multiple radiation-based tests. Even if such practices were feasible, the dynamic nature of the human body renders this approach far from ideal. 
Consequently, there is a growing need for models that require less ground truth data, aligning more closely with clinical realities. Our model is tailored to meet this demand.

In the pretraining stage, the proposed method relies solely on noisy data. Only in the fine-tuning stage, both noisy and ground truth clean data become necessary to support a supervised training paradigm. In this context, the proposed model can leverage the noisy data to encode the structural dependence across various anatomical regions and thus potentially demands less ground truth data to learn meaningful representations.

To empirically validate this distinctive feature, SwinIR is studied since it exhibits superior denoising performance as demonstrated in Subsection C. We randomly select different numbers of patients for model training and then evaluate the models performances under the varying data configurations. As shown in Fig. \ref{depend}, as the number of patients in the training data increases, both SSIM and RMSE metrics show noticeable improvement for both SwinIR and SwinIR+LoMAE models. 
Generally speaking, the curve of SwinIR+LoMAE method lies above SwinIR's in SSIM and under in RMSE. To simplify notation, the number following each method indicates the quantity of patients with ground-truth data used for model training. For instance, 'LoMAE2' denotes that the model was trained with data from only two patients with clean images. It's particularly noteworthy that even with only two patients' worth of clean data, SwinIR+LoMAE consistently outperforms SwinIR, clearly demonstrating its advantage. Furthermore, t-tests are performed between LoMAE2 and SwinIR methods. The LoMAE2 method shows significant improvement ($p<0.05$) over SwinIR8 ($p=6.07\times 10^{-3}$), SwinIR6 ($p=5.99\times 10^{-3}$), SwinIR4 ($p=2.56 \times 10^{-3}$), and SwinIR2 ($p=1.22 \times 10^{-3}$).

\begin{figure*}
\centering
\includegraphics[width=\textwidth]{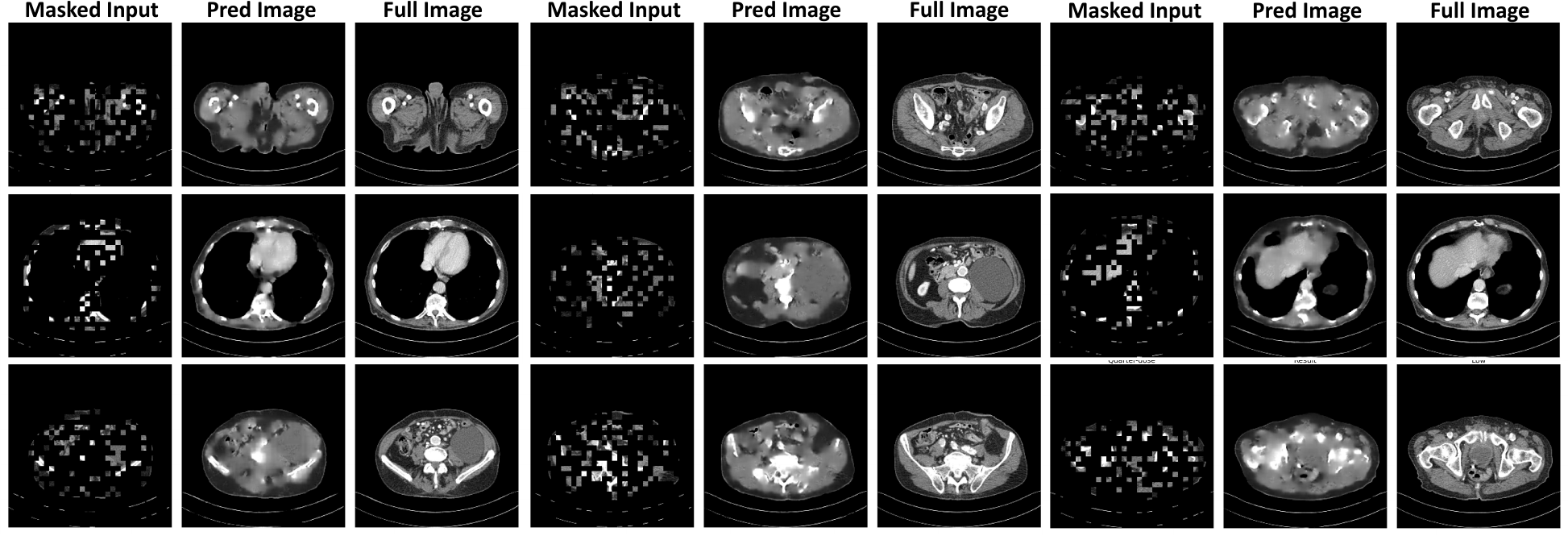}
\caption{The MAE reconstruction results on the testing cases. The display window for the gray images is [-160,240] \textit{HU}.} 
\label{mae_result}
\end{figure*}

\begin{figure*}
\centering
\includegraphics[width=\linewidth]{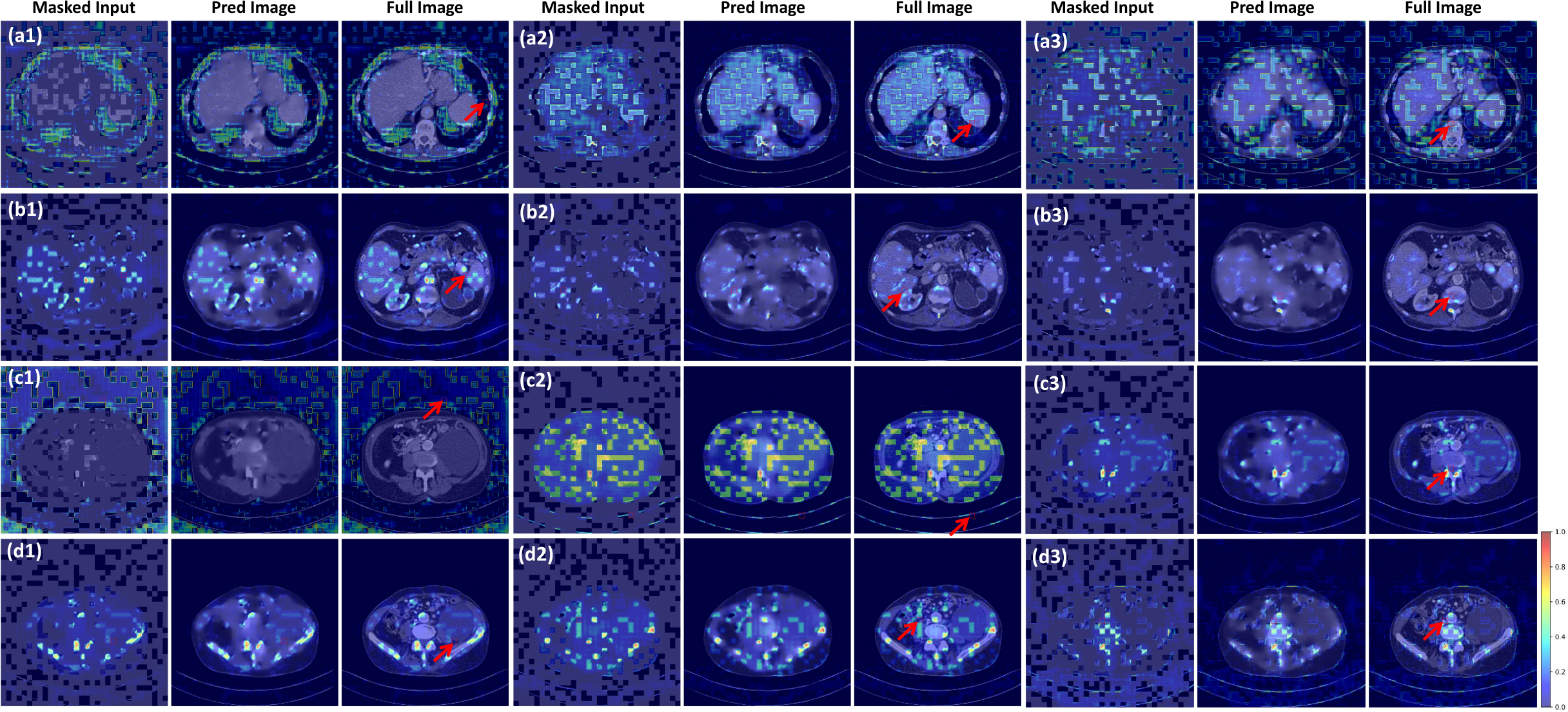}
\caption{The MAE-GradCAM results from four representative slices. Each row represents one slice, while the three columns denote three different locations. The red arrows indicate the interested $8\times8$ patch areas for the MAE-GradCAM. It is better to view this image in a zoomed version. } 
\label{cam}
\end{figure*}
Moreover, results of all slices in Fig. \ref{scatter} provide a visual representation of these findings. It illustrates that the majority of curves of the SwinIR+LoMAE are better than that of SwinIR, which indicates that the SwinIR+LoMAE method can greatly reduce its reliance on ground truth data. This attribute makes it a highly practical choice for clinical scenarios where noisy data is abundant while labeled data remains scarce.

\begin{figure*}
\centering
\includegraphics[width=\linewidth]{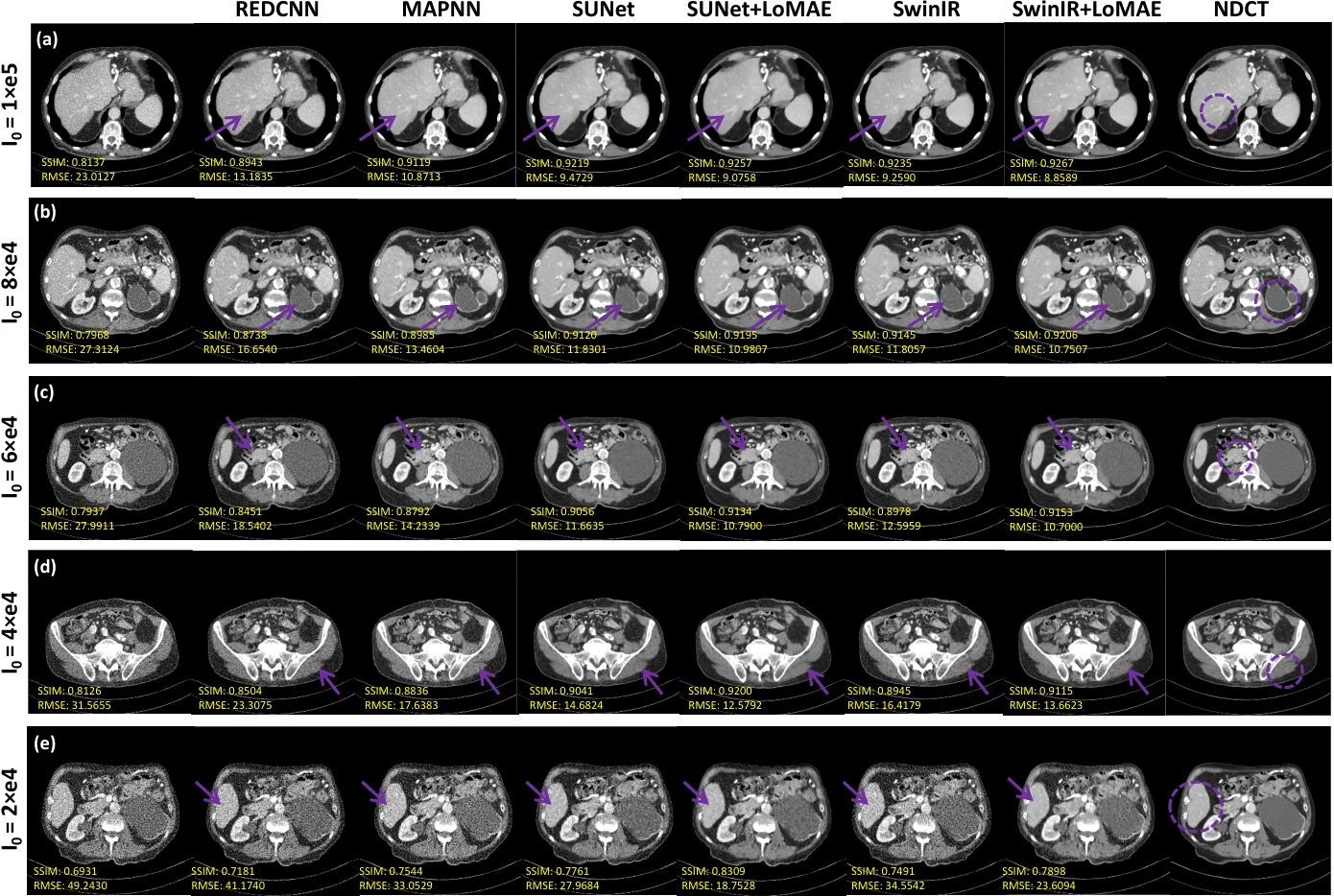}
\caption{Representative testing results of the same trained models under different noise levels (x-ray intensity). The display window is [-160,240] \textit{HU}.} 
\label{levels}
\end{figure*}

\subsection{MAE Interpretation}
In the pretraining phase, the MAE adeptly learns the mapping from corrupted images to their original counterparts. The results, as depicted in Figure \ref{mae_result}, demonstrate the remarkable capability of the MAE, indicating that even when only 25\% of the original image is available, it can effectively reconstruct the entire legitimate image with remarkable accuracy. While minor details are missing, the overall structural integrity and anatomical regions remain largely preserved post-pretraining, underscoring the efficacy of this pretraining process.


Furthermore, with the proposed MAE-GradCAM, some latent behaviors of the MAE pretraining can be decoded. As shown from the representative slices in Fig. \ref{cam}, although the exact causal relationship of the patch generation process is inevident, several common behaviors can be unveiled:
\begin{itemize}
    \item The MAE can clearly discriminate the foreground body tissue and the background areas. In all exemplary slices, the attention is distinctly focused either on the foreground or background. 
    \item The saliency map with similar weights indicates similar tissues or organs, suggesting that MAE potentially possesses remarkable semantic segmentation capability. For example in Fig. \ref{cam} (b1, b2, c3, d1, d3), the attention on the liver and kidney is consistently distributed across the entire organs. 
    \item The completeness of the missing background is typically highly affected by other existing background patches. For instance in Fig. \ref{cam} (c1), the missing background patches rely on the presence of other background areas. 
    \item The boundaries between low-intensity and high-intensity are predominantly determined by the high-intensity organs and partially determined by the low-intensity areas. For example in Fig. \ref{cam} (b1, b2, b3, c3, d1, d2, d3), the intersected area is primarily determined by the high-intensity area like aorta, rib, and ilium, with a less influence from low-intensity areas like liver, kidney, and spleen. 
\end{itemize}

\subsection{Model Robustness and  Generalizability}
Deep learning models excel at learning complicated latent functions. However, their performance often hinges on the assumption that testing data follows the same distribution as the training data. Controversially, the data in the real world is much more diverse and complex with many unpredictable perturbations, thus, posing a severe challenge to a model's robustness and generalizability. In our case, a clinically favorable denoising model should deliver constant good denoising performance across various noise distributions. Naturally, we are motivated to study the robustness and generalizability of the proposed method. Specifically, we evaluate how the quarter dose-trained model performs over different noise distributions. 


\begin{figure}
\centering
\includegraphics[width=\linewidth]{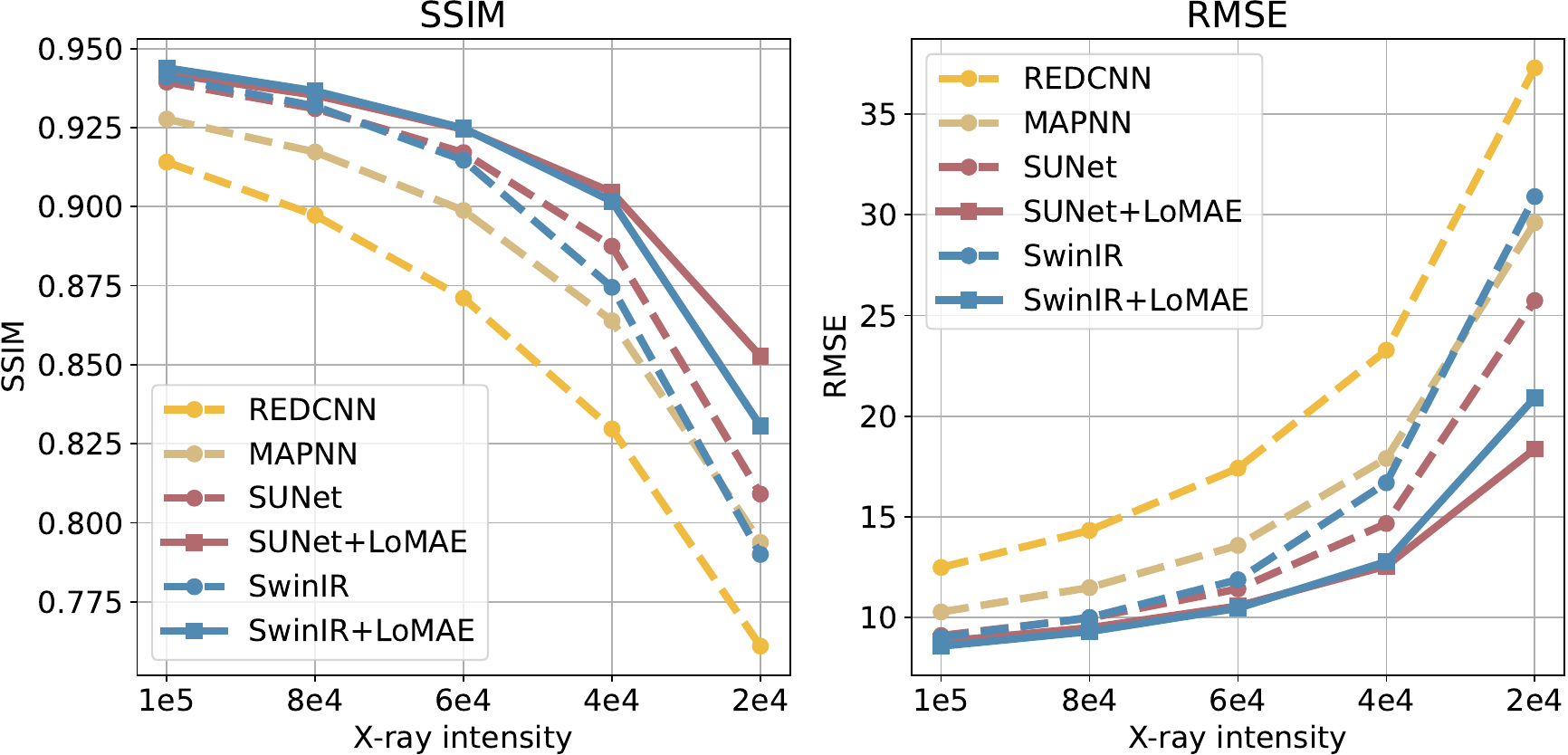}
\caption{The quantitative curves of different methods under different noise levels.} 
\label{noise}
\end{figure}

\begin{figure}
\centering
\includegraphics[width=\linewidth]{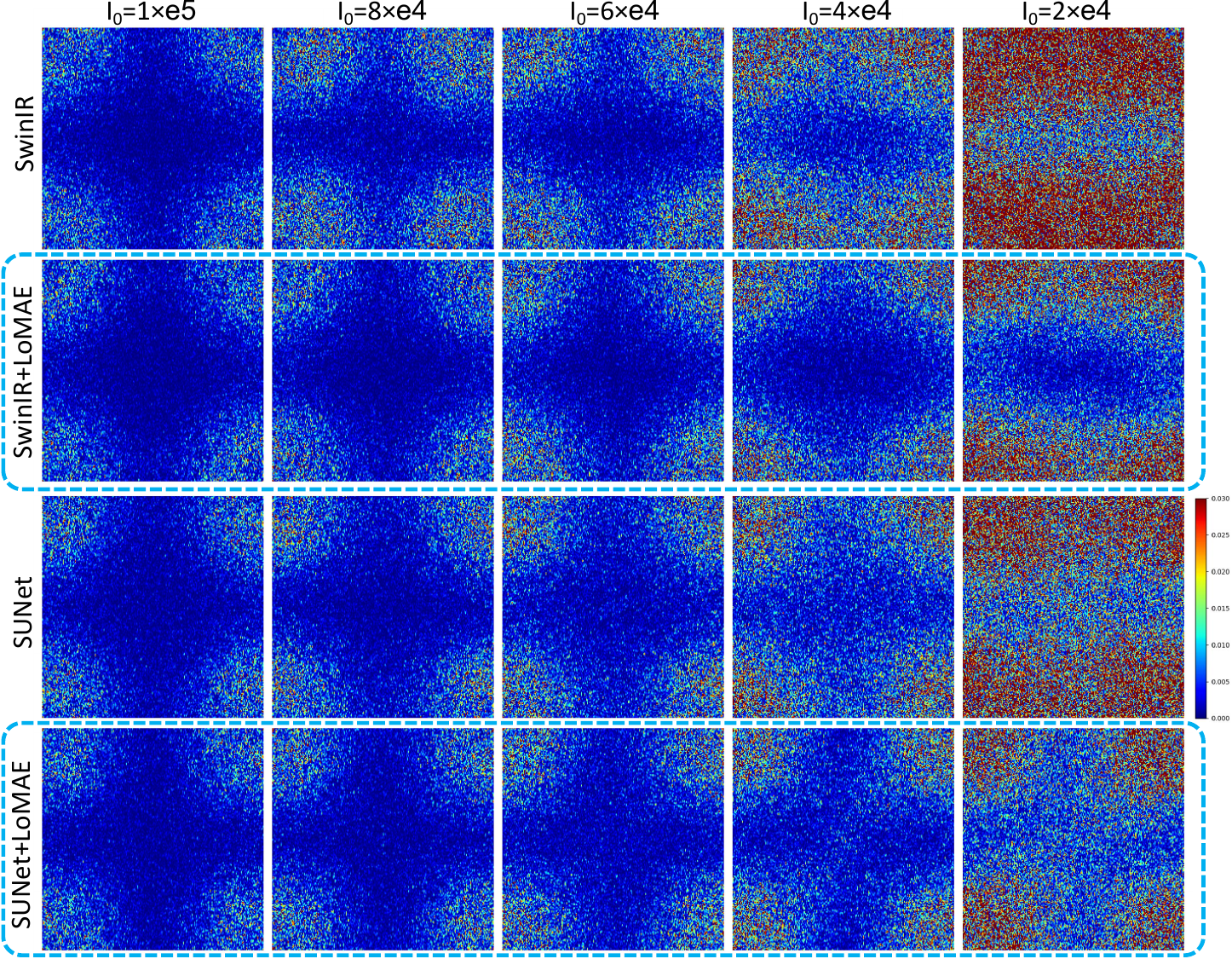}
\caption{Comparison of the NPS maps SwinIR and SUNet with respect to the corresponding LoMAE methods under different noise levels. The image center of each map represents low frequency structural noise while the edges and corners indicate the high frequency noise such as Poisson and Gaussian. } 
\label{nps}
\end{figure}
\begin{figure}
\centering
\includegraphics[width=\linewidth]{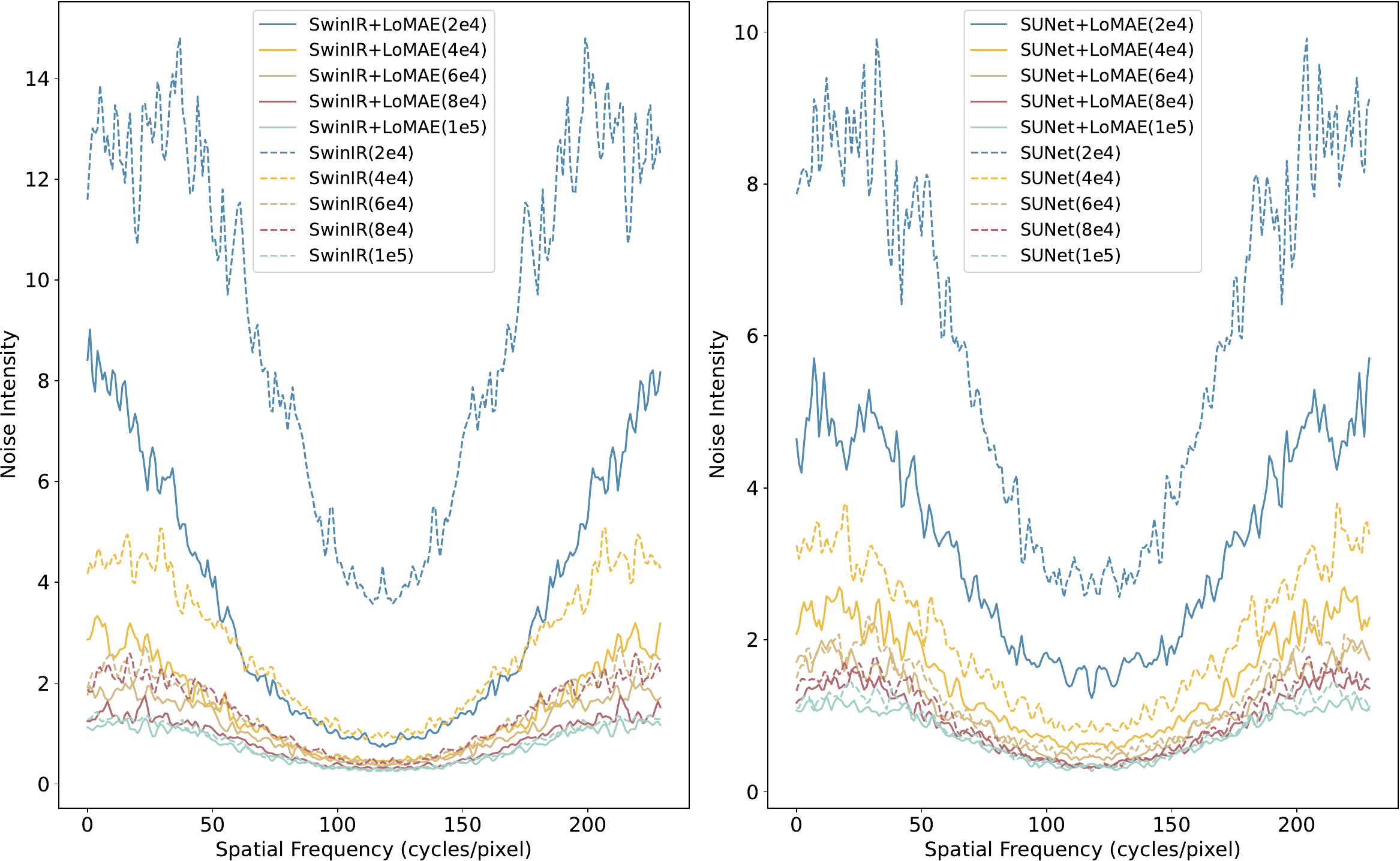}
\caption{The cumulative outcomes of the NPS curves along the vertical axis. The left figure displays the results of SwinIR and SwinIR+LoMAE while the right figure indicates SUNet and SUNet+LoMAE.} 
\label{nps1d}
\end{figure}

Experiments are conducted on the \textit{Simulated Dataset} where a variety of noise settings is performed. The noise level is mainly determined by the x-ray intensities (dose level), \emph{e.g.}, $1\times 10^5$, $8\times 10^4$, $6\times 10^4$, $4\times 10^4$, and $2\times 10^4$, which correspond to $10\%$, $8\%$, $6\%$, $4\%$, and $2\%$ doses, respectively.
As shown in Fig. \ref{levels}, the studied methods exhibit denoising capabilities across different noise levels. Unsurprisingly, all methods' performance drops when the noise level increases, which is expected since all models are only trained on 25\% dose images. Notably, our method exhibits a slower decline in denoising capability and constantly delivers the best results across all tests as evidenced by both qualitative results in Fig. \ref{levels} and quantitative curves in Fig. \ref{noise}. For example in Fig. \ref{levels}(a), the LoMAE methods capture the hepatic vein structure with high fidelity. In Fig. \ref{levels}(b) where there exists a solid cystic mass in the left kidney, the LoMAEs generate smoother interior structures. Next in Fig. \ref{levels}(c), LoMAE can deliver the most accurate colon image with minimal distortion. Finally, Fig. \ref{levels}(d) and Fig. \ref{levels}(e) illustrate that our models render clearer muscle and kidney tissues compared to others. Particularly at the lowest intensity of $\mathrm{I}_0 = 2\times 10^4$, all other methods fail to work but our method, especially the SwinIR+LoMAE, can still suppress the major noises and retain the key structures. 

Moreover, the noise power spectrum (NPS) maps of SwinIR and SUNet are visualized along with the corresponding LoMAE-enhanced variants in Fig. \ref{nps}. These maps indicate that the LoMAE methods are subject to less noise across different noise frequencies, especially on high-frequency noise. The horizontally summed results in Fig. \ref{nps1d} further underscore the superior noise reduction and structure preservation capabilities of LoMAE methods compared to the raw approaches. Particularly, the low-frequency noise (middle of Fig. \ref{nps1d} on x axis) is most pronounced in SwinIR ($2 \times 10^4$) and SUNet ($2 \times 10^4$), indicating significant structural losses in these cases. Controversially, the corresponding low-frequency noise is significantly lower in the LoMAE-enhanced methods. 

\begin{figure}
\centering
\includegraphics[width=\linewidth]{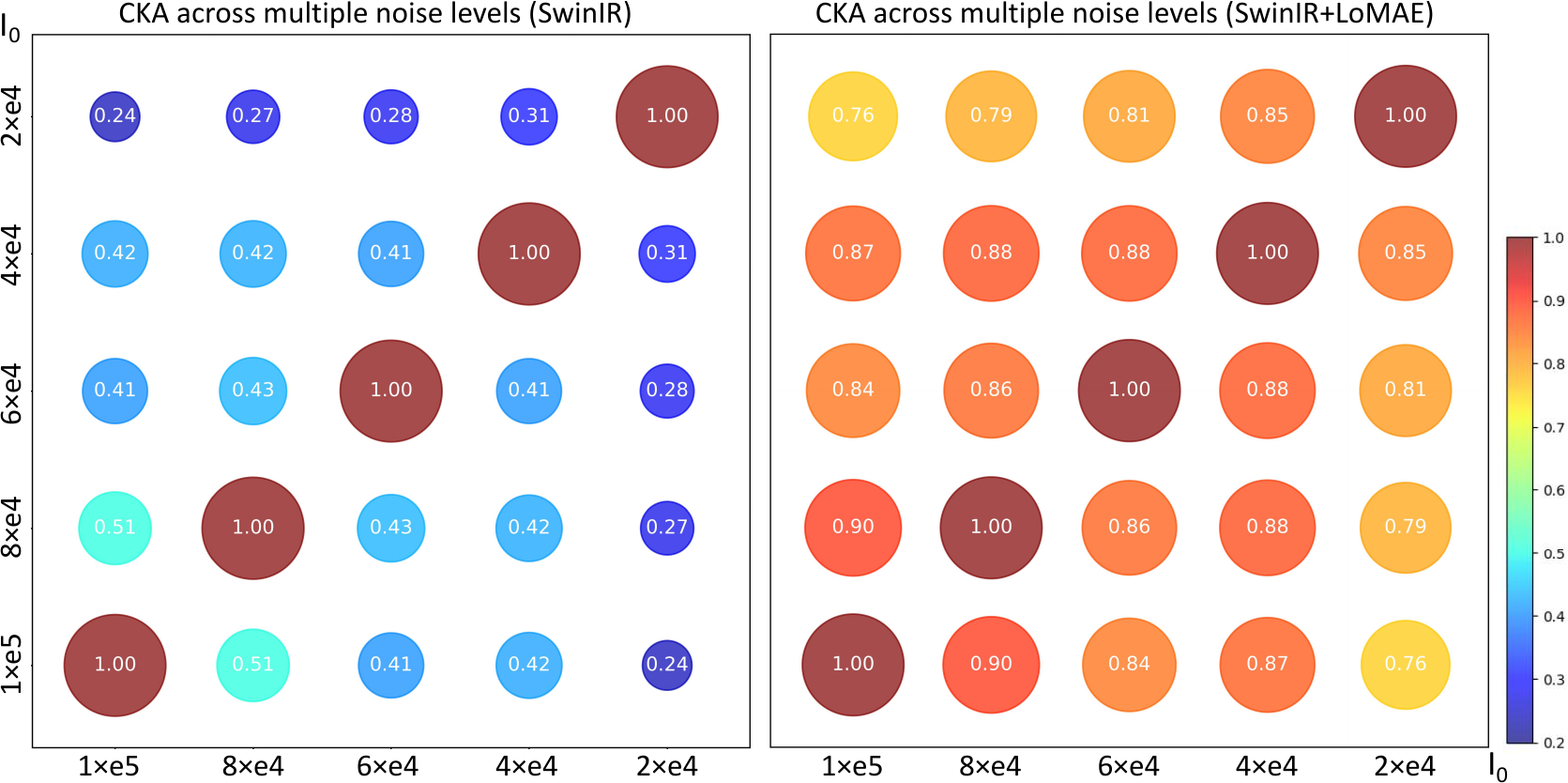}
\caption{The CKA results on SwinIR and SwinIR+LoMAE. CKAs are calculated between the last-layer feature maps of input images with different noise levels. } 
\label{cka}
\end{figure}

To further demonstrate the generalizability of the proposed LoMAE, centered kernel alignment (CKA) \cite{cortes2012algorithms,raghu2021vision} is performed on the feature maps of SwinIR and LoMAE. Specifically, let's consider $\mathbf{X}_1 \in R^{m\times d}$ and $\mathbf{X}_2 \in R^{m\times d}$ as two token layers to compute the CKA. The Gram matrices is denoted as $\mathbf{K}=\mathbf{X}_1\mathbf{X}_1^\top$ and $\mathbf{L}=\mathbf{X}_2\mathbf{X}_2^\top$, the CKA is expressed as:
\begin{equation}
    \mathrm{CKA}(\mathbf{K},\mathbf{L}) = \frac{\mathrm{HSIC}(\mathbf{K},\mathbf{L})}{\sqrt{\mathrm{HSIC(\mathbf{K},\mathbf{K})}\mathrm{HSIC}(\mathbf{L},\mathbf{L})}}.
\end{equation}
Here, $\mathrm{HSIC}(\cdot)$ represents the Hilbert-Schhmidt Independence Criterion, which is calculated as below:
\begin{equation}
    \mathrm{HSIC}(\mathbf{K},\mathbf{L}) = \frac{vec(\mathbf{K}') \cdot vec(\mathbf{L}')}{(m-1)^2},
\end{equation}
where $\mathbf{K}'=\mathbf{H}\mathbf{K}\mathbf{H}$ and $\mathbf{L}'=\mathbf{H}\mathbf{L}\mathbf{H}$ are the centered Gram metrics of $\mathbf{K}$ and $\mathbf{L}$, respectively. $\mathbf{H}=\mathbf{I}_n - \frac{1}{n}\mathbf{1}_n \mathbf{1}_n^\top$ is the centering metric. And $vec(\cdot)$ denotes the vectorization of a matrix. 

In this analysis, CKA is calculated among the feature maps over various noise settings. Higher CKA indicates higher generalizability across multiple noise levels. As shown in Fig. \ref{cka}, the LoMAE method demonstrates significantly higher CKAs than raw SwinIR. Particularly, the CKA between $2\times 10^4$ and $1\times 10^5$ is 0.24 with SwinIR, but 0.76 with LoMAE. Therefore, it can be concluded that the proposed LoMAE possesses strong robustness and generalizability across different dose levels.

\subsection{Ablation Studies}

\textit{On the impact of base model:} In this study, we primarily utilize SwinIR and SUNet for LOMAE applications. Nevertheless, this strategy can also be applied to non-swin based transformers. As an example, the LoMAE is tested on Restormer \cite{zamir2022restormer}, another SOTA transformer model for denoising. The original Restormer achieves SSIM and RMSE values of $0.9552\pm{0.0010}$ and $7.4543\pm{0.1084}$, respectively. With Restormer+LoMAE, the SSIM and RMSE improve to $0.9582\pm{0.0009}$ and $6.9881\pm{0.1022}$, respectively. Those results further confirm the efficacy of the LoMAE on raw transformer models. 

\textit{On the impact of the front-to-end shortcut:} Front-to-end shortcut is absent in LoMAE pretaining but utilized in finetuning to ensure information consistency. To evaluate its usefulness, we remove the shortcut in finetuning as a comparison. Quantitative results on SUNet show that without the shortcut, the SSIM and RMSE are $0.9571\pm{0.0012}$ and $7.0627\pm{0.1000}$, which are inferior to the results of $0.9601\pm{0.0008}$ and $6.8352\pm{0.0900}$ obtained with the shortcut. 

\section{Conclusion}

This paper explored the novel application of Masked Autoencoder in LDCT denoising, marking the pioneering use of MAE as a self-pretraining strategy in low-level denoising tasks. Specifically, we proposed a LoMAE design and developed an MAE-GradCAM method to interpret the latent learning behavior. Experimental results on the AAPM and simulated dataset underscored LoMAE's exceptional ability to enhance noise removal and structural preservation, producing qualitatively and quantitatively superior CT images. Remarkably, LoMAE proved effective in semi-supervised contexts, significantly reducing reliance on ground truth data. Furthermore, our findings demonstrated LoMAE's high robustness and generalizability across a spectrum of noise levels, paving the way for potential applications in various medical image modalities. Our future research endeavors will focus on further exploring LoMAE's applicability in these domains.




\bibliographystyle{ieeetr}
\bibliography{reference.bib}

\end{document}